\documentclass[preprint,12pt,superscriptaddress]{revtex4}  

\usepackage{graphics}
\usepackage[usenames]{color}
\usepackage{epsfig}

\def\cp{$CP$\/}

\def\mevm{~MeV/$c^2$\/}
\def\mevp{~MeV/$c$\/}
\def\meve{~MeV}
\def\gevm{~GeV/$c^2$\/}

\def\geve{~GeV}

\def\ra{\!\rightarrow\!}

\def\bbar{\overline{B}{}^{\,0}}

\def\bsdsds{$B^0_s\ra D^{(*)+}_s D^{(*)-}_s$}
\def\bsdsstdsst{$B^0_s\ra D^{*+}_s D^{*-}_s$}
\def\bsdsstds{$B^0_s\ra D^{*\pm}_s D^{\mp}_s$}
\def\bsdspi{$B^0_s\ra D^{(*)-}_s \pi^+$}
\def\bdsd{$B^0\ra D^{(*)+}_s D^-$}

\def\bs{B^{}_s}
\def\bsst{B^{*}_s}
\def\bsbar{\overline{B}{}^{}_s}
\def\bsbarst{\overline{B}{}^{\,*}_s}
\def\fl{f^{}_L}

\def\mbc{M^{}_{\rm bc}}
\def\de{\Delta E}
\def\qq{$q\bar{q}$}

\def\dgs{\Delta\Gamma^{}_s}
\def\dgcp{\Delta\Gamma^{CP}_s}
\def\gs{\Gamma^{}_s}
\def\kstz{\overline{K}{}^{\,*0}}
\def\kstp{K^{*+}}
\def\qqbar{$q\bar{q}$}

\def\simge{\mathrel{%
   \rlap{\raise 0.511ex \hbox{$>$}}{\lower 0.511ex \hbox{$\sim$}}}}
\def\simle{\mathrel{
   \rlap{\raise 0.511ex \hbox{$<$}}{\lower 0.511ex \hbox{$\sim$}}}}

\begin{document}



\title{\large
\begin{flushright}
{\normalsize KEK Preprint 2012-20} \\
\vspace*{-0.17in}
{\normalsize UCHEP-12-11} 
\end{flushright}
\boldmath{
Precise measurement of the branching fractions for \bsdsds\ 
and first measurement of the $D^{*+}_s D^{*-}_s$ polarization using 
$e^+e^-$ collisions 
}}

\affiliation{Budker Institute of Nuclear Physics SB RAS and Novosibirsk State University, Novosibirsk 630090}
\affiliation{Faculty of Mathematics and Physics, Charles University, Prague}
\affiliation{University of Cincinnati, Cincinnati, Ohio 45221}
\affiliation{Hanyang University, Seoul}
\affiliation{University of Hawaii, Honolulu, Hawaii 96822}
\affiliation{High Energy Accelerator Research Organization (KEK), Tsukuba}
\affiliation{Indian Institute of Technology Guwahati, Guwahati}
\affiliation{Indian Institute of Technology Madras, Madras}
\affiliation{Institute of High Energy Physics, Chinese Academy of Sciences, Beijing}
\affiliation{Institute of High Energy Physics, Vienna}
\affiliation{Institute of High Energy Physics, Protvino}
\affiliation{Institute for Theoretical and Experimental Physics, Moscow}
\affiliation{J. Stefan Institute, Ljubljana}
\affiliation{Kanagawa University, Yokohama}
\affiliation{Institut f\"ur Experimentelle Kernphysik, Karlsruher Institut f\"ur Technologie, Karlsruhe}
\affiliation{Korea Institute of Science and Technology Information, Daejeon}
\affiliation{Korea University, Seoul}
\affiliation{Kyungpook National University, Taegu}
\affiliation{\'Ecole Polytechnique F\'ed\'erale de Lausanne (EPFL), Lausanne}
\affiliation{Faculty of Mathematics and Physics, University of Ljubljana, Ljubljana}
\affiliation{Luther College, Decorah, Iowa 52101}
\affiliation{University of Maribor, Maribor}
\affiliation{Max-Planck-Institut f\"ur Physik, M\"unchen}
\affiliation{University of Melbourne, School of Physics, Victoria 3010}
\affiliation{Graduate School of Science, Nagoya University, Nagoya}
\affiliation{Kobayashi-Maskawa Institute, Nagoya University, Nagoya}
\affiliation{Nara Women's University, Nara}
\affiliation{National Central University, Chung-li}
\affiliation{National United University, Miao Li}
\affiliation{Department of Physics, National Taiwan University, Taipei}
\affiliation{H. Niewodniczanski Institute of Nuclear Physics, Krakow}
\affiliation{Nippon Dental University, Niigata}
\affiliation{Niigata University, Niigata}
\affiliation{University of Nova Gorica, Nova Gorica}
\affiliation{Osaka City University, Osaka}
\affiliation{Pacific Northwest National Laboratory, Richland, Washington 99352}
\affiliation{Research Center for Electron Photon Science, Tohoku University, Sendai}
\affiliation{University of Science and Technology of China, Hefei}
\affiliation{Seoul National University, Seoul}
\affiliation{Sungkyunkwan University, Suwon}
\affiliation{School of Physics, University of Sydney, NSW 2006}
\affiliation{Tata Institute of Fundamental Research, Mumbai}
\affiliation{Excellence Cluster Universe, Technische Universit\"at M\"unchen, Garching}
\affiliation{Tohoku Gakuin University, Tagajo}
\affiliation{Tohoku University, Sendai}
\affiliation{Department of Physics, University of Tokyo, Tokyo}
\affiliation{Tokyo Institute of Technology, Tokyo}
\affiliation{Tokyo Metropolitan University, Tokyo}
\affiliation{Tokyo University of Agriculture and Technology, Tokyo}
\affiliation{CNP, Virginia Polytechnic Institute and State University, Blacksburg, Virginia 24061}
\affiliation{Yamagata University, Yamagata}
\affiliation{Yonsei University, Seoul}
\author{S.~Esen}\affiliation{University of Cincinnati, Cincinnati, Ohio 45221} 
\author{A.~J.~Schwartz}\affiliation{University of Cincinnati, Cincinnati, Ohio 45221} 
  \author{H.~Aihara}\affiliation{Department of Physics, University of Tokyo, Tokyo} 
  \author{D.~M.~Asner}\affiliation{Pacific Northwest National Laboratory, Richland, Washington 99352} 
  \author{T.~Aushev}\affiliation{Institute for Theoretical and Experimental Physics, Moscow} 
  \author{A.~M.~Bakich}\affiliation{School of Physics, University of Sydney, NSW 2006} 
  \author{K.~Belous}\affiliation{Institute of High Energy Physics, Protvino} 
  \author{B.~Bhuyan}\affiliation{Indian Institute of Technology Guwahati, Guwahati} 
  \author{A.~Bozek}\affiliation{H. Niewodniczanski Institute of Nuclear Physics, Krakow} 
  \author{M.~Bra\v{c}ko}\affiliation{University of Maribor, Maribor}\affiliation{J. Stefan Institute, Ljubljana} 
% ...
\collaboration{The Belle Collaboration}

\begin{abstract}
We have made a precise measurement of the absolute branching
fractions of \bsdsds\ decays using 121.4~fb$^{-1}$ of data recorded 
by the Belle experiment running at the $\Upsilon(5S)$ resonance. 
The results are 
${\cal B}(B^0_s\ra D^+_s D^-_s)=(0.58\,^{+0.11}_{-0.09}\,\pm0.13)\%$,
${\cal B}(B^0_s\ra D^{*\pm}_s D^{\mp}_s)=(1.76\,^{+0.23}_{-0.22}\,\pm 0.40)\%$, 
and
${\cal B}(B^0_s\ra D^{*+}_s D^{*-}_s)=(1.98\,^{+0.33}_{-0.31}\,^{+0.52}_{-0.50})\%$;
the sum is 
${\cal B}(B^0_s\ra D^{(*)+}_s D^{(*)-}_s)=(4.32\,^{+0.42}_{-0.39}\,^{+1.04}_{-1.03})\%$.
Assuming \bsdsds\ saturates decays to \cp-even final states, the 
branching fraction constrains the ratio $\dgs/\cos\phi^{}_{12}$, where 
$\dgs$ is the difference in widths between the two $\bs$-$\bsbar$ 
mass eigenstates, and $\phi^{}_{12}$ is the \cp-violating phase in 
$\bs$-$\bsbar$ mixing. 
We also measure for the first time the longitudinal polarization fraction 
of \bsdsstdsst; the result is $0.06\,^{+0.18}_{-0.17}\,\pm 0.03$.
\end{abstract}


\maketitle

Decays of $\bs$ mesons help elucidate the weak 
Cabibbo-Kobayashi-Maskawa structure of the Standard Model (SM).
$\bs$ decays can be studied at $e^+e^-$ colliders by 
running at the $\Upsilon(5S)$ resonance, which decays
to $B^{(*)}_s\overline{B}{}^{(*)}_s$ pairs. We have used this
method previously~\cite{esen10} to study \bsdsds\ decays
using 23.6~fb$^{-1}$ of data. Here we substantially improve
this measurement using 121.4~fb$^{-1}$ of data. In addition
to the five-times-larger data set, there are other improvements 
to the analysis: the data have been fully reprocessed using 
reconstruction algorithms with higher efficiency for 
$\pi^0$'s and low momentum tracks; we use larger control 
samples to evaluate systematic uncertainties; and we take 
background probability density functions directly from 
data rather than from simulation.
We also make the first measurement of the fraction of 
longitudinal polarization ($f^{}_L$) of \bsdsstdsst.

As in our previous study, we reconstruct the final states 
$D^+_sD^-_s$, 
$D^{*+}_sD^-_s\!+\!D^{*-}_sD^+_s$ ($\equiv D^{*\pm}_sD^\mp_s$),
and $D^{*+}_sD^{*-}_s$.
These are expected to be mostly \cp-even, and 
their partial widths are expected to dominate the difference 
in widths between the two $\bs$-$\bsbar$ \cp\ eigenstates,
$\dgcp$~\cite{Aleksan}. This parameter equals $\dgs/\cos\phi^{}_{12}$, 
where $\dgs$ is the decay width difference between the mass 
eigenstates, and $\phi^{}_{12}\!=\!{\rm arg}(-M_{12}/\Gamma_{12})$, 
where $M^{}_{12}$ and $\Gamma^{}_{12}$ are the off-diagonal elements 
of the $\bs$-$\bsbar$ mass and decay matrices~\cite{Dunietz}. 
The phase $\phi^{}_{12}$ is the \cp-violating phase in $\bs$-$\bsbar$ 
mixing. Thus the branching fraction gives a constraint in the 
$\dgs$-$\phi^{}_{12}$ parameter space. Both parameters can receive 
contributions from new physics (NP)~\cite{Nierste,new_physics}. 
Previous constraints on $\dgs$ and NP contributions to $\phi^{}_{12}$ 
were obtained from a time-dependent angular analysis of 
$B^{}_s\ra J/\psi\,\phi$ decays~\cite{previous_D0,previous_CDF,LHCb_deltagamma}.
A constraint on $\phi^{}_{12}$ can be derived from the \cp\ 
asymmetry measured in $B^{}_s$ semileptonic decays~\cite{HFAG}.
 
At the $\Upsilon(5S)$ resonance, the $e^+e^-\ra b\bar{b}$ cross 
section is measured to be $\sigma^{}_{b\bar{b}}=0.340\pm 0.016$~nb,
and the fraction of $\Upsilon(5S)$ decays producing $\bs$ mesons 
is $f^{}_s=0.172\,\pm 0.030$~\cite{alexey+remi_full}. 
Thus the total number of $\bs\bsbar$ pairs is 
$N^{}_{\bs\bsbar} = 
(121.4~{\rm fb}^{-1})\cdot\sigma^{}_{b\bar{b}}\cdot f^{}_s = 
(7.11\pm 1.30)\times 10^6$.
Three production modes are kinematically allowed: $\bs\bsbar$, 
$\bs\bsbarst$ or $\bsst\bsbar$, and $\bsst\bsbarst$. 
The production fractions ($f^{}_{B^{(*)}_s\overline{B}{}^{\,(*)}_s}$)
for the latter two are $0.073\,\pm 0.014$ and $0.870\,\pm 0.017$, 
respectively~\cite{remi_full}. 
The $\bsst$ decays via $\bsst\ra\bs\gamma$, and the $\gamma$ 
is not reconstructed. 

The Belle detector running at the KEKB $e^+e^-$ collider~\cite{kekb} 
is described in Ref.~\cite{belle_detector}. For charged hadron 
identification, a likelihood ratio is formed based on $dE/dx$ 
measured in the central tracker and the response of aerogel
threshold \u{C}erenkov counters and time-of-flight 
scintillation counters. 
A likelihood requirement is used to identify charged kaons and 
pions. This requirement is 86\% efficient for $K^\pm$ and has 
a $\pi^\pm$ misidentification rate of~8\%.

We reconstruct $B^0_s\ra D^+_s D^-_s$, $D^{*\pm}_s D^{\mp}_s$, 
and $D^{*+}_s D^{*-}_s$ decays in which 
$D^+_s\ra \phi\pi^+$,
$K^0_S\,K^+$,
$\kstz K^+$,
$\phi\rho^+$,
$K^0_S\,\kstp$, and 
$\kstz \kstp$~\cite{charge-conjugates}.
Neutral $K^0_S$ candidates 
are reconstructed from $\pi^+\pi^-$ pairs having an invariant mass 
within 10\mevm\ of the nominal $K^0_S$ mass~\cite{pdg} and satisfying 
momentum-dependent vertex requirements.
Charged tracks are required to originate from near the $e^+e^-$ 
interaction region and, with the exception of tracks from $K^0_S$ 
decays, have a momentum $p>\!100$\mevp. 
Neutral $K^{*0}$ (charged $K^{*+}$) candidates are reconstructed 
from a $K^+\pi^-$ ($K^0_S\,\pi^+$) pair 
having an invariant mass within 50\mevm\ of $m^{}_{K^{*}}$. 
Candidate $\phi$ mesons are reconstructed from $K^+K^-$  pairs having 
an invariant mass within 12\mevm\  of $m^{}_{\phi}$. Charged $\rho^+$ 
candidates are reconstructed from $\pi^+\pi^0$ pairs 
having an invariant mass within 100\mevm\  of $m^{}_{\rho^+}$.
The $\pi^0$ candidates are reconstructed from $\gamma\gamma$ 
pairs having an invariant mass within 15\mevm\ of $m^{}_{\pi^0}$,
and with each $\gamma$ having an energy $E^{}_\gamma>\!100$\meve. 

The invariant mass windows used for the reconstructed 
$D^+_s$ candidate (denoted $\widetilde{D}_s^+$) are:
$\pm 10$\mevm\ ($\sim\!3\sigma$) for the three final 
states containing $K^*$ candidates,
$\pm 20$\mevm\ ($2.8\sigma$) for $\phi\rho^+$,
and $\pm 15$\mevm\ ($\sim\!4\sigma$) for the remaining two 
modes. For the three vector-pseudoscalar final states we 
require $|\cos\theta^{}_{\rm hel}|>0.20$, where 
$\theta^{}_{\rm hel}$ is the angle between the momentum of 
the charged daughter of the vector particle and the direction 
opposite the $\widetilde{D}_s^+$ 
momentum, evaluated in the rest frame of the vector particle. 

We combine $D^+_s$ candidates with photon candidates 
to reconstruct $D^{*+}_s\ra D^+_s\gamma$ decays.
We require 
$E^{}_\gamma\!>\!50$\meve\ in the 
$e^+e^-$ center-of-mass system,
and that the energy deposited in the central $3\times 3$ 
array of cells of the electromagnetic cluster exceeds
85\% of that deposited in the central $5\times 5$ array of cells.
The mass difference 
$M^{}_{\widetilde{D}_s^+\gamma} - M^{}_{\widetilde{D}_s^+}$ 
is required to be within 12.0\mevm\ of the nominal value.
This requirement and also that of the $\widetilde{D}_s^+$ mass windows are 
determined by optimizing a figure-of-merit $S/\sqrt{S+B}$, where $S$ is 
the expected signal based on Monte Carlo (MC) simulation and $B$ is the 
background estimated from either MC simulation or $D^+_s$ mass sideband data.

We select $B^0_s\ra D^+_s D^-_s$, $D^{*\pm}_s D^{\mp}_s$,  and 
$D^{*+}_s D^{*-}_s$ decays using two quantities evaluated in the 
center-of-mass frame: 
the beam-energy-constrained mass $\mbc=\sqrt{E^2_{\rm beam} - p^2_B}$, 
and the energy difference $\de= E^{}_B-E^{}_{\rm beam}$, where 
$p^{}_B$ and $E^{}_B$ are the reconstructed momentum and energy 
of the $B^0_s$ candidate, and $E_{\rm beam}$ is the beam energy. 
We determine signal yields by fitting events satisfying 
$5.25\mbox{\gevm}<\mbc <5.45$\gevm\ and $-0.15\mbox{\geve}<\de <0.10$\geve.
Because the $\gamma$ from $\bsst\ra\bs\gamma$ is not reconstructed,
the modes $\Upsilon(5S)\ra\bs\bsbar$, $\bs\bsbarst$ and 
$\bsst\bsbarst$ are well-separated in $\mbc$ and $\de$.
We expect only small contributions from $\bs\bsbar$ 
and $\bs\bsbarst$ events and fix these contributions 
relative to $\bsst\bsbarst$ according to our measurement 
using $B_s^0 \to D^-_s \pi^+$ decays~\cite{remi_full}. 
We quote fitted signal yields from $\bsst\bsbarst$ only 
and use these to determine the branching fractions.

Approximately half of selected events have multiple \bsdsds\ candidates.
These typically arise from photons produced via $\pi^0\ra\gamma\gamma$
that are wrongly assigned as $D_s^*$ daughters. For these events 
we select the candidate that minimizes the quantity
$$
\frac{1}{(2+N)}\,\Biggl\{
\sum_{D^{}_s} \left[ \frac{M^{}_{\widetilde{D}_s} - 
M^{}_{D_s}}{\sigma^{}_M}\right]^2 +
\sum_{D_s^{*}} \left[ 
\frac{\Delta\widetilde{M} - \Delta M }
{\sigma^{}_{\Delta M}}\right]^2\Biggr\}\,,
$$
where 
$\Delta\widetilde{M}=M^{}_{\widetilde{D}_s^+\gamma} - M^{}_{\widetilde{D}_s^+}$ 
and $\Delta M =M^{}_{D^{*+}_s}-M^{}_{D^+_s}$.
The summations run over the two $D^+_s$ daughters and the 
$N$ ($=\!0,1,2$) $D^{*+}_s$ daughters 
of a $B^0_s$ candidate. The mean masses $M^{}_{D^{(*)}_s}$ 
and widths $\sigma^{}_M$ and $\sigma^{}_{\Delta M}$ are obtained 
from MC simulation and calibrated for data-MC differences using 
a large $B^0\ra D^{(*)+}_sD^-$ control sample from $\Upsilon(4S)$ 
data. According to the simulation, this criterion selects 
the correct candidate 83\%, 73\%, and 69\% of the time
for $D^+_s D^-_s$, $D^{*\pm}_s D^{\mp}_s$, and 
$D^{*+}_s D^{*-}_s$ states, respectively.

We reject background from $e^+e^-\ra q\bar{q}~(q=u,d,s,c)$  
events using a  Fisher discriminant based on a set of modified Fox-Wolfram 
moments~\cite{KSFW}. This discriminant distinguishes jet-like \qq\ events 
from more spherical $B^{}_{(s)}\overline{B}{}^{}_{(s)}$ events. 
With this discriminant we calculate likelihoods 
$\mathcal{L}_{s}$ and $\mathcal{L}_{q\overline q}$ 
for an event assuming the event is signal or $q\overline{q}$ 
background; we then require 
$\mathcal{L}_{s}/(\mathcal{L}_{s}+\mathcal{L}_{q\overline q})\!>\!0.20$.
This selection is 93\% efficient for signal events and removes 
more than 62\% of \qqbar\ background events.

The remaining background consists of 
$\Upsilon(5S)\ra B^{(*)}_s\overline{B}{}^{(*)}_s\ra D^+_s X$,
$\Upsilon(5S)\ra B\overline{B}X$ ($b\bar{b}$ hadronizes 
to $B^0,\,\bbar$, or $B^\pm$), and
$B^{}_s\ra D^\pm_{sJ}(2317)D^{(*)}_s$, $D^\pm_{sJ}(2460)D^{(*)}_s$, 
or $D^\pm_{s}D^\mp_s\pi^0$. 
The last three processes peak at negative $\de$, and their 
yields are estimated to be small using analogous 
$B^{}_d\ra D^\pm_{sJ}D^{(*)}$ branching fractions. We thus 
consider them only when evaluating systematic uncertainty 
due to backgrounds. All selection criteria are finalized 
before looking at events in the signal region. 

We measure signal yields by performing a two-dimensional 
unbinned maximum-likelihood
fit to the $\mbc$-$\de$ distributions. For each sample, 
we include probability density functions (PDFs) for signal 
and $q\bar{q}$, $B^{(*)}_s\overline{B}{}^{(*)}_s\ra D^+_s X$, 
and $\Upsilon(5S)\ra BBX$ backgrounds. As the
backgrounds have similar $\mbc$ and $\de$ shapes, 
we use a single PDF for them, taken to be an ARGUS function~\cite{ARGUS} 
for $\mbc$ and a first-order Chebyshev function for $\de$. 
The two parameters of the Chebyshev function are taken from data in which 
one of the $D^+_s$ candidates is required to be within the mass sideband.

The signal PDFs have three components: correctly reconstructed (CR) decays; 
``wrong combination'' (WC) decays in which a non-signal track or $\gamma$ is 
included in place of a true daughter track or $\gamma$; and ``cross-feed'' (CF) 
decays in which a $D^{*\pm}_s D^{\mp}_s$ ($D^{*+}_s D^{*-}_s$) is reconstructed 
as a $D^+_s D^-_s$ ($D^+_s D^-_s$ or $D^{*\pm}_s D^{\mp}_s$), or a $D^+_s D^-_s$ 
($D^{*\pm}_s D^{\mp}_s$) is reconstructed as a $D^{*\pm}_s D^{\mp}_s$ 
or $D^{*+}_s D^{*-}_s$ ($D^{*+}_s D^{*-}_s$). 
In the former case, the $\gamma$ from
$D^{*+}_s\ra D^+_s\gamma$ is lost and $\de$ is shifted down
by $100\!-\!150$\meve; this is called ``CF-down.'' In the latter
case, an extraneous $\gamma$ is included and $\de$ is shifted 
up by a similar amount; this is called ``CF-up.'' In both 
cases $\mbc$ remains almost unchanged. 
 
All signal shape parameters are taken from MC simulation and 
calibrated using \bsdspi\ and \bdsd\ decays. The CR PDF is taken
to be a Gaussian for $\mbc$ and a double Gaussian with common 
mean for $\de$. 
The CF and WC PDFs consist of sums of Gaussians and a Chebyshev 
function for $\de$, and Gaussians and either a Novosibirsk 
function~\cite{Novosibirsk} or a Crystal Ball function~\cite{Crystal_ball}
for $\mbc$. The fractions of WC and CF-down events are taken from the 
simulation.
The fractions of CF-up events are floated as they are difficult 
to simulate accurately (i.e., many $B_s^0$ partial widths are 
unmeasured). As the CF-down fractions are fixed, the separate
$D^+_s D^-_s$, $D^{*\pm}_s D^{\mp}_s$, and $D^{*+}_s D^{*-}_s$
samples are fitted simultaneously.

\begin{figure}
\hbox{\hskip0.6in
\epsfig{file=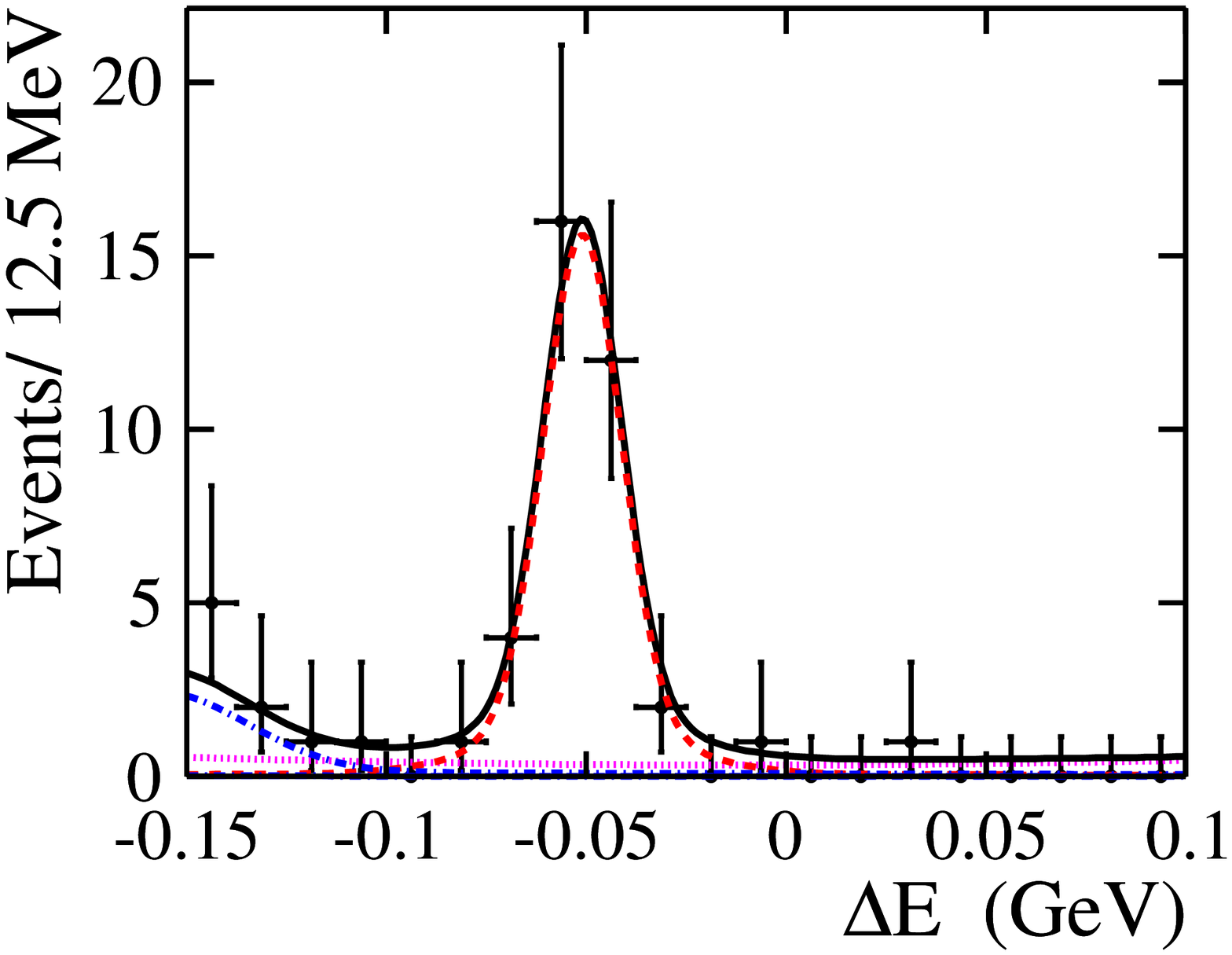,width=2.5in}
\hskip0.10in
\epsfig{file=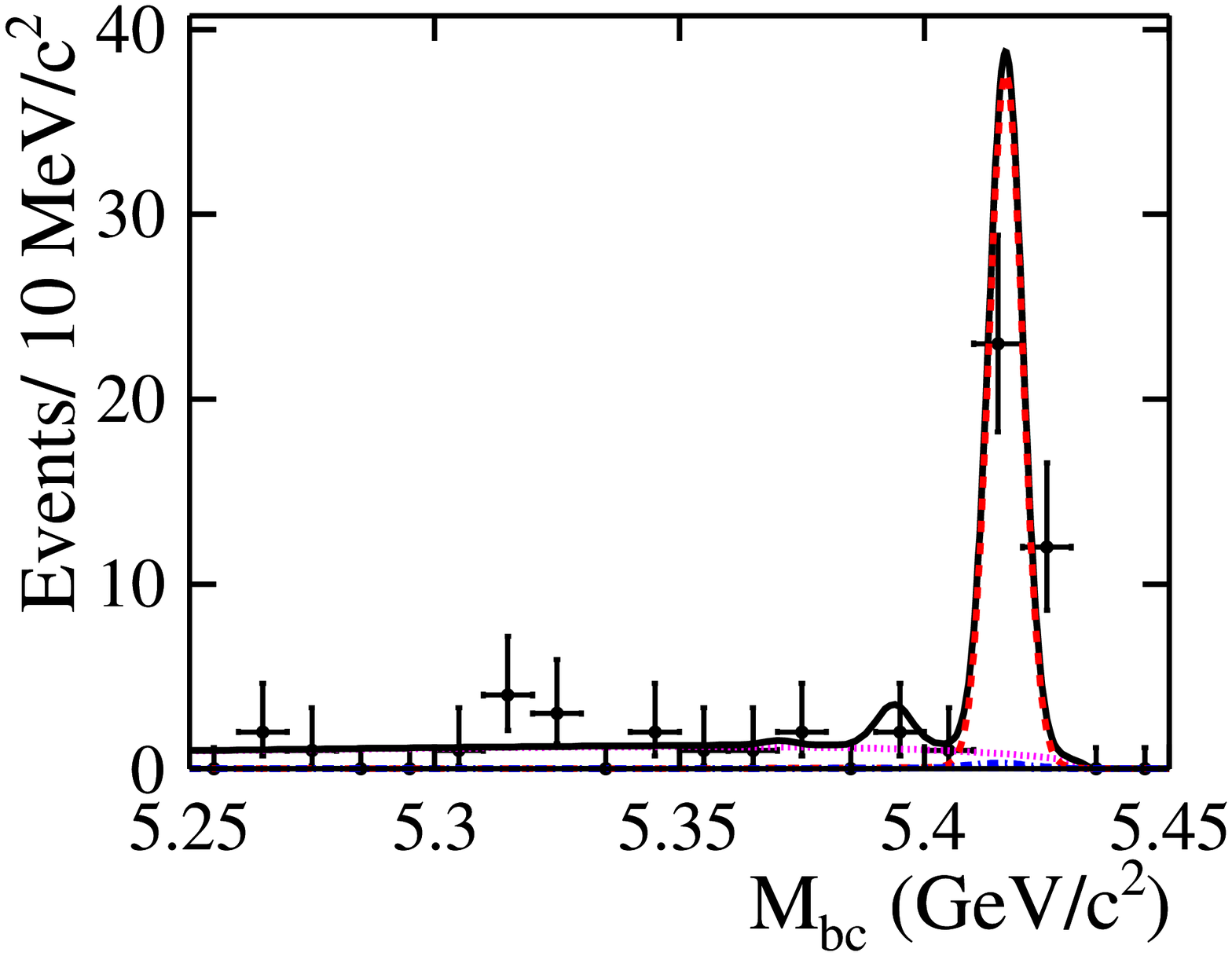,width=2.5in}
}
\hbox{\hskip0.6in
\epsfig{file=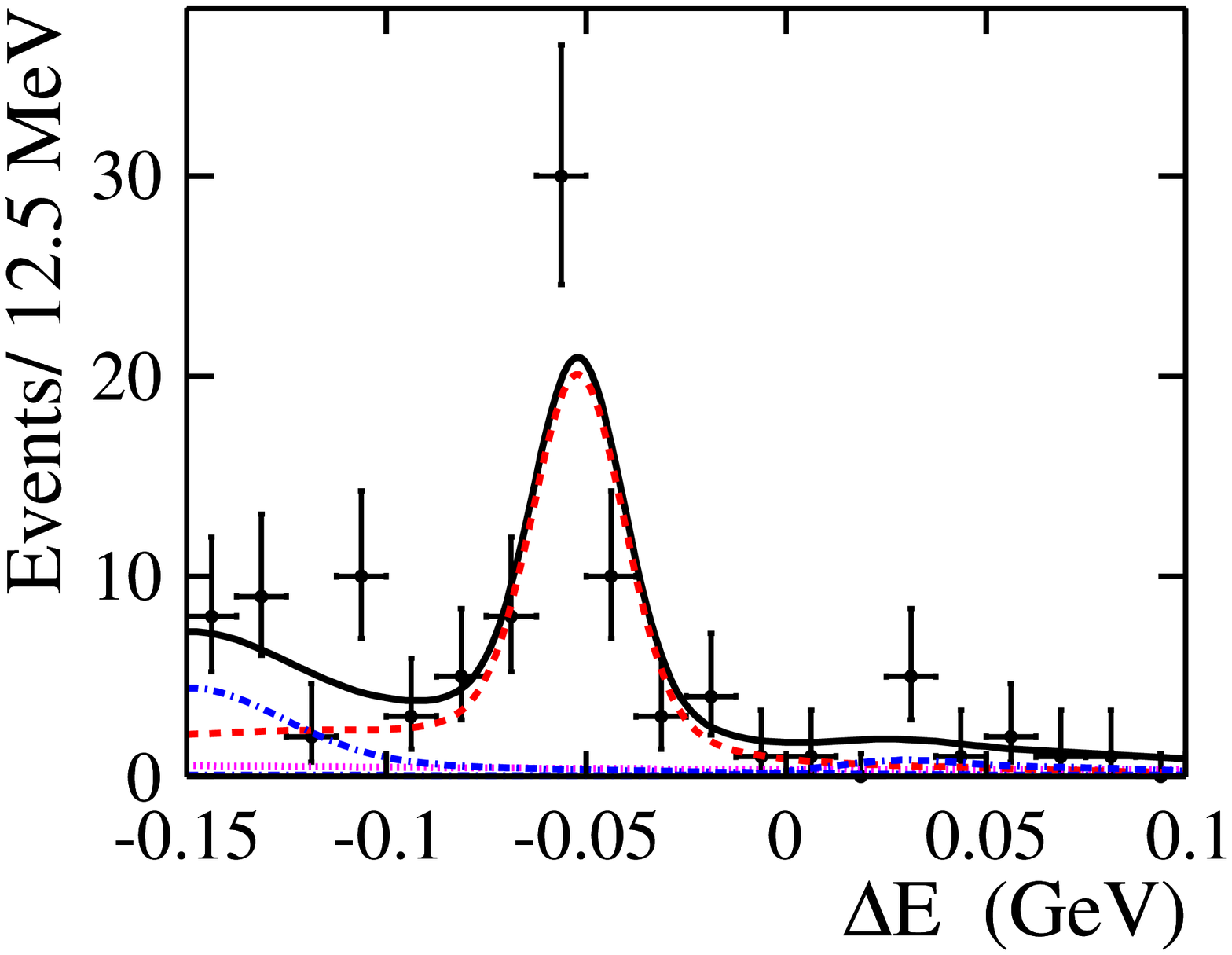,width=2.5in}
\hskip0.10in
\epsfig{file=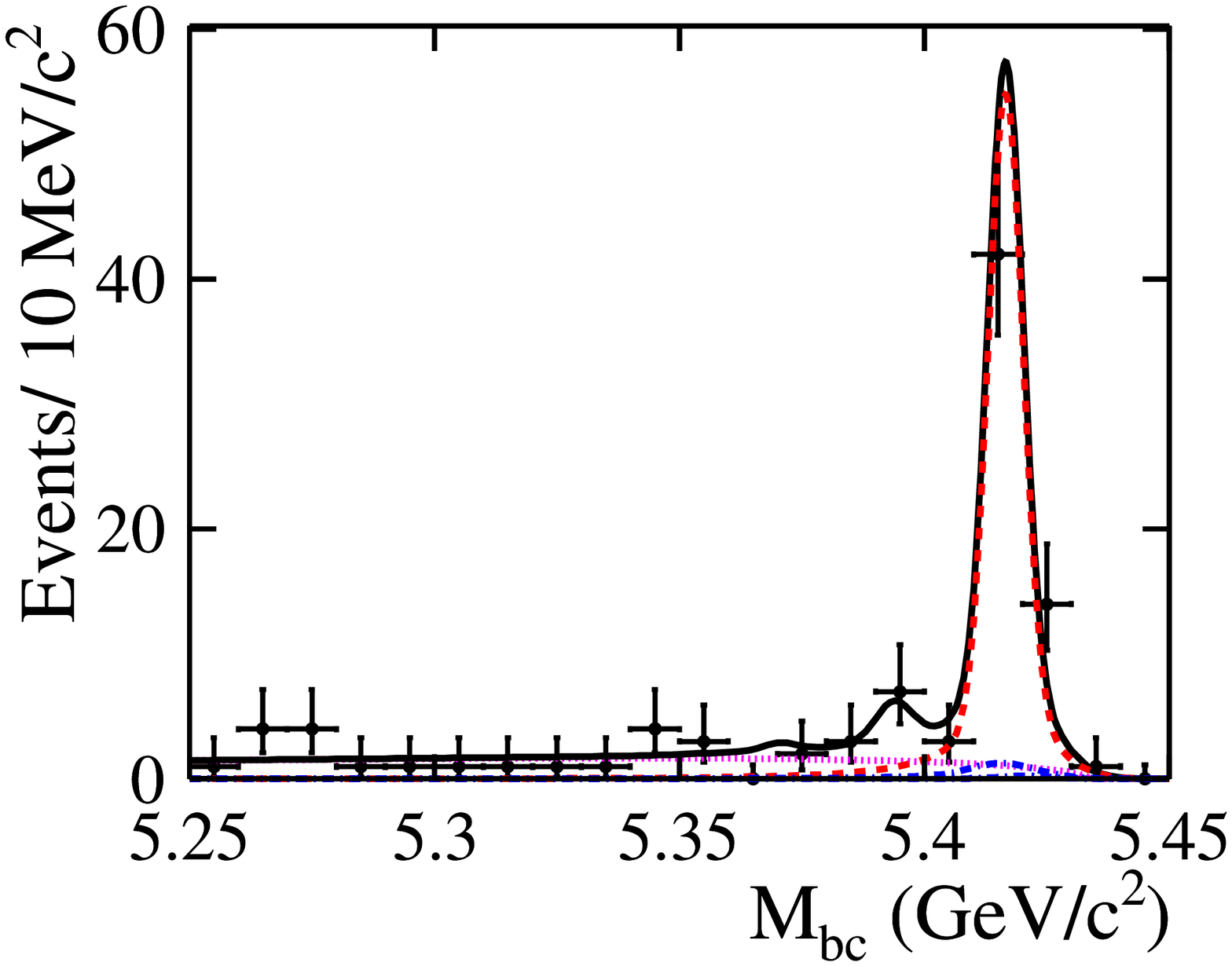,width=2.5in}
}
\hbox{\hskip0.6in
\epsfig{file=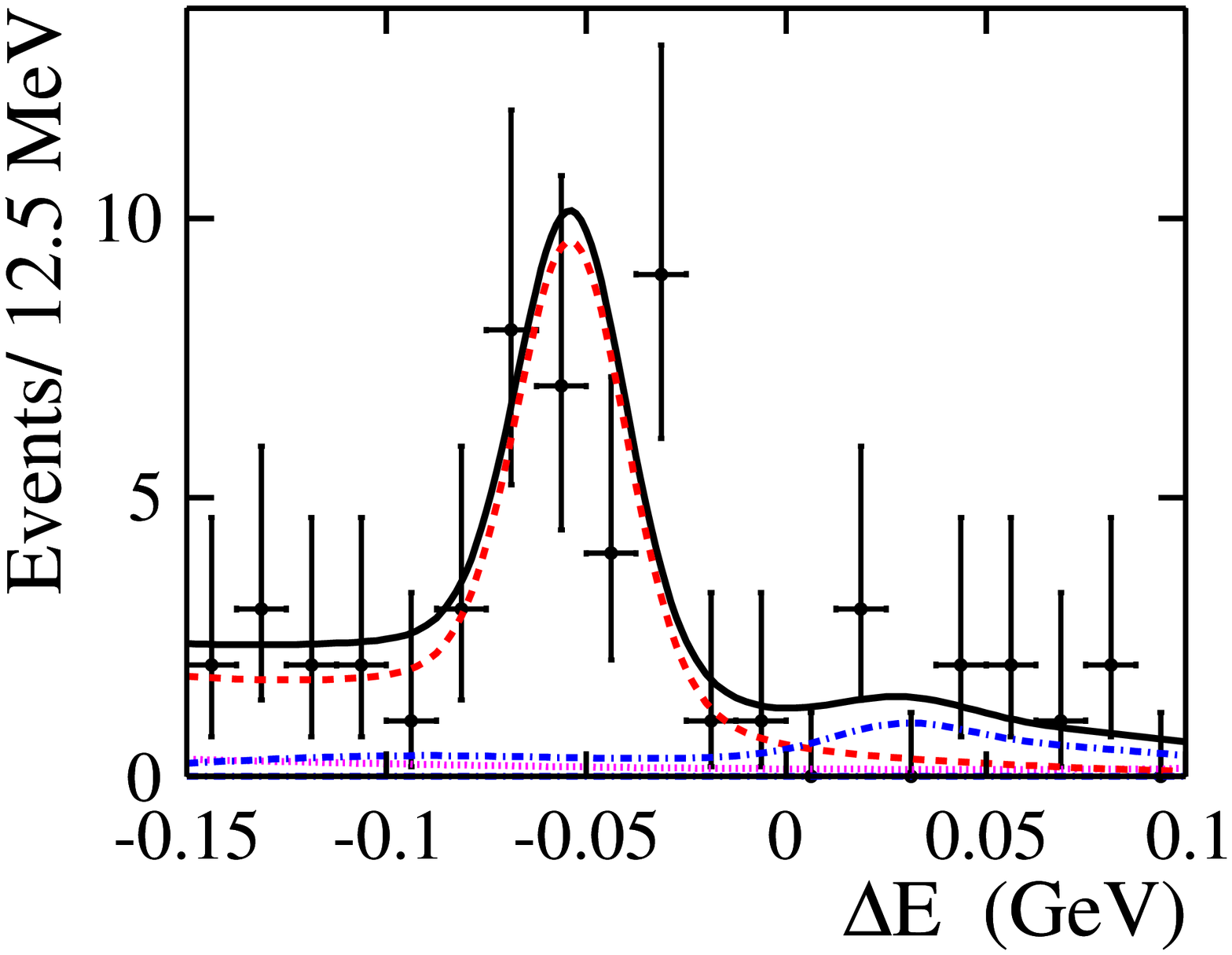,width=2.5in}
\hskip0.10in
\epsfig{file=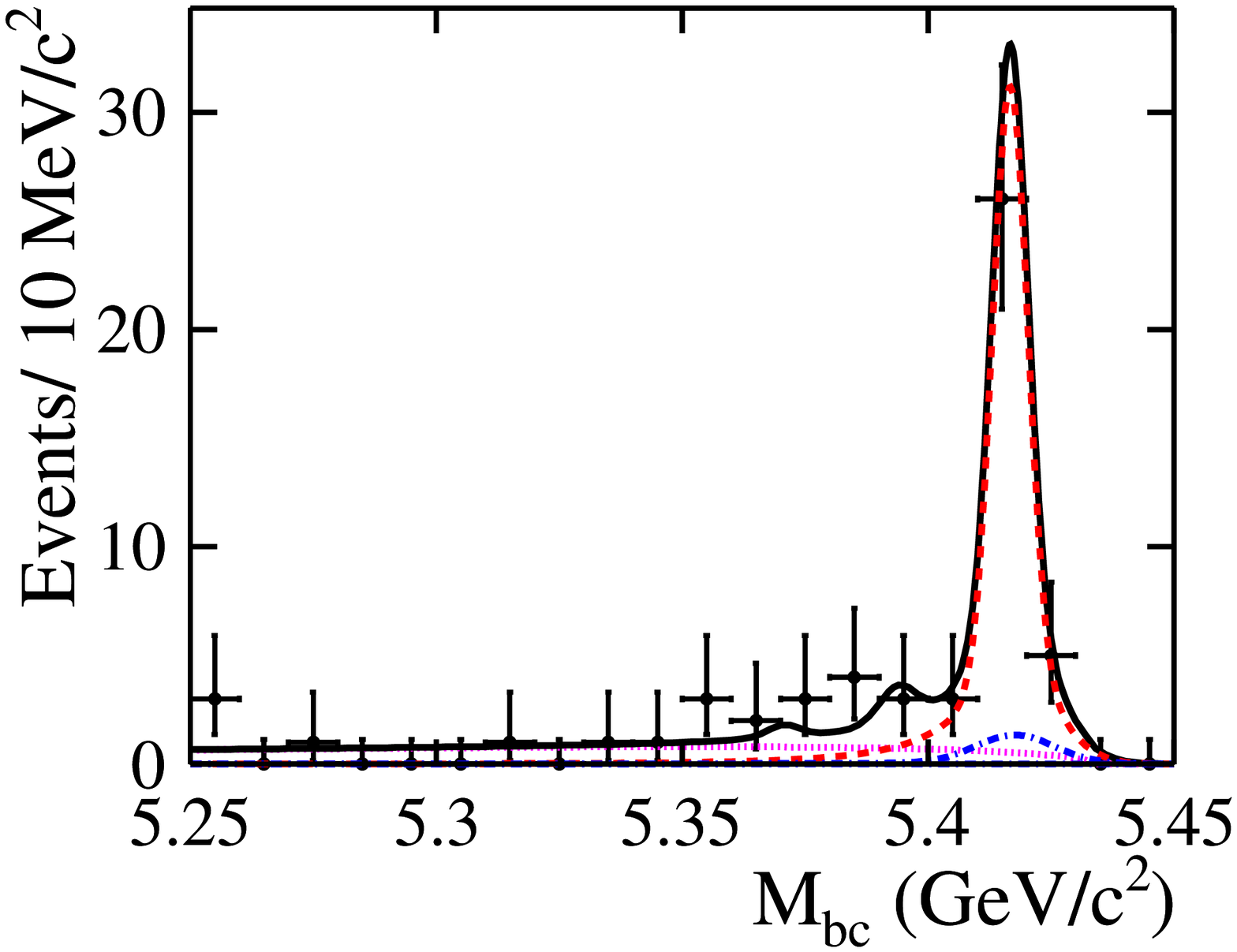,width=2.5in}
}
\caption{
$\de$ fit projections for events satisfying
$M^{}_{\rm bc} \in [5.41, 5.43]$~GeV/$c^2$, and
$\mbc$ fit projections for events satisfying
$\Delta E \in [-0.08, -0.02]$~GeV.
The top row shows $B^0_s\ra D^+_s D^-_s$; 
the middle row shows $B^0_s\ra D^{*\pm}_s D^{\mp}_s$; and 
the bottom row shows $B^0_s\ra D^{*+}_s D^{*-}_s$. 
The red dashed curves show CR+WC signal; 
the blue dash-dotted curves show CF;
the magenta dotted curves show background; 
and the black solid curves show the total.} 
\label{fig:fit_results}
\end{figure}

The projections of the fit are shown in Fig.~\ref{fig:fit_results},
and the fitted signal yields are listed in Table~\ref{tab:fit_results}. 
The branching fraction for channel $i$ is calculated as
${\cal B}^{}_i = Y^{}_i/(\varepsilon^i_{MC}\cdot N^{}_{\bs\bsbar}
\cdot f^{}_{B^*_s\overline{B}{}^{\,*}_s}\cdot 2)$, where $Y^{}_i$ is the 
fitted CR yield, and $\varepsilon^i_{MC}$ is the MC signal efficiency 
with intermediate branching fractions~\cite{pdg} included.
The efficiencies $\varepsilon^i_{MC}$ include small 
correction factors to account for differences between MC and 
data for kaon identification. Inserting all values gives the 
branching fractions listed in Table~\ref{tab:fit_results}.
These results have similar precision as other recent 
measurements~\cite{CDF_deltagamma} and are in 
agreement with theoretical predictions~\cite{Sanda,Hou}.
The statistical significance is calculated
as $\sqrt{-2\ln(\mathcal{L}_0 / \mathcal{L}_{\mathrm{max}})}$, 
where $\mathcal{L}_0$ and $\mathcal{L}_{\mathrm{max}}$ are the  
values of the likelihood function when the signal yield $Y^{}_i$
is fixed to zero and when it is floated, respectively. 
We include systematic uncertainties (discussed below) in 
the significance by smearing the likelihood function by 
a Gaussian having a width equal to the total systematic 
error related to the signal yield.

The systematic errors are listed in Table~\ref{tab:syst_errors}. 
The error due to PDF shapes is evaluated by varying shape parameters 
by $\pm 1\sigma$. The errors for the fixed WC and CF-down  
fractions are evaluated by repeating the fit with each fixed fraction 
varied by $\pm 20$\%. 
Those fractions that are correlated (e.g., WC for $D_s^*D^+_s$ and 
$D_s^{*+}D^{*-}_s$, which is due to reconstructing extraneous photons)
are varied together in the ratio predicted from MC simulation.
The systematic errors due to $q\bar{q}$ suppression 
and the best candidate selection are evaluated using
control samples of $B^0_s\ra D^-_s\pi^+$ and $B^0\ra D^{(*)+}_s D^-$,
respectively. These errors are taken as the change in
the branching fractions when the criteria are applied.
The uncertainties due to $\pi^\pm/K^\pm$ identification 
and tracking efficiency are obtained from 
$D^{*+}\ra D^0\pi^+\ra K^-\pi^+\pi^+$ decays; 
these are $\sim$\,1\% and 0.35\% per track, respectively.
Significant uncertainties arise from the $\Upsilon(5S)\ra\bsst\bsbarst$ 
and $D^+_s$ branching fractions, which are external factors. 
We take the $D^{*+}_sD^{*-}_s$ polarization $\fl$ 
for this measurement to be the 
well-measured value from the analogous decay 
$B^0_d\ra D^{*+}_s D^{*-}$: $0.52\pm\,0.05$~\cite{pdg}. 
The systematic error is taken as the change in ${\cal B}$ 
when $f^{}_L$ is varied over a wide range: from $2\sigma$ 
higher than 0.52 down to the (low) central value 
we measure below.

\begin{table}[ptbh]
\caption{$\bsst\bsbarst$ CR signal yield ($Y$) and efficiency ($\varepsilon$),
including intermediate branching fractions, and resulting
branching fraction (${\cal B}$) and signal significance ($S$), 
including systematic errors. The first errors listed are 
statistical; the others are systematic. The last error for 
the sum is due to external factors ($\Upsilon(5S)\ra\bsst\bsbarst$ 
and $D_s^+$ branching fractions). }
\begin{center}
\renewcommand{\arraystretch}{1.0}
\begin{tabular*}{0.75\textwidth}
{@{\extracolsep{\fill}} l | c c c c }
\hline\hline
Mode & $Y$ & $\varepsilon$ & ${\cal B}$ & $S$ \\
 & (events) & ($\times 10^{-4}$) & (\%) &   \\
\hline
$D^+_s D^-_s$ & 
 $33.1_{-5.4}^{+6.0}$ & 4.72 &  $0.58\,^{+0.11}_{-0.09}\,\pm0.13$ & 11.5\\  [1.0ex] 
$D^{*\pm}_s D^{\mp}_s$ & 
$44.5_{-5.5}^{+5.8}$ & 2.08 &  $1.76\,^{+0.23}_{-0.22}\,\pm 0.40$ & 10.1\\  [1.0ex] 
$D^{*}_s D^{*}_s$ & 
$24.4_{-3.8}^{+4.1}$ & 1.01 &  $1.98\,^{+0.33}_{-0.31}\,^{+0.52}_{-0.50}$ & 7.8\\  [1.0ex] 

\hline
Sum & 
$102.0_{-8.6}^{+9.3}$ &  &  $4.32\,^{+0.42}_{-0.39}\,^{+0.56}_{-0.54}\,\pm0.88$ & \\  [0.5ex] 
\hline\hline
\end{tabular*} 
\end{center}
\label{tab:fit_results}
\end{table}

\begin{table}[ptbh]
\caption{\label{tab:syst_errors}
Systematic errors (\%). Those listed in the top section affect the 
signal yield and thus the signal significance.}
\centering
\renewcommand{\arraystretch}{0.9}
\begin{tabular*}{0.75\textwidth}{@{\extracolsep{\fill}} l  | c c c c  c c @{\extracolsep{\fill}} }
\hline
\hline
Source & 
\multicolumn{2}{c}{$D^+_sD^-_s$} & 
\multicolumn{2}{c}{$D^*_s D^{}_s$} & 
\multicolumn{2}{c}{$D^{*+}_s D^{*-}_s$} \\
\hline
 & $+\sigma$ & $-\sigma$ & $+\sigma$ & $-\sigma$ & $+\sigma$ & $-\sigma$ \\
\hline
Signal PDF shape   & 2.7 &2.2   & 2.2& 2.4& 5.1& 3.8 \\
Bckgrnd PDF shape & 1.5 &1.3   & 1.3& 1.4& 2.9& 2.8 \\
WC + CF fraction   & 0.5 & 0.5 & 4.7 & 4.5 & 11.0 & 9.7 \\
$q\bar{q}$ suppression
    & 	3.1 & 0.0	& 0.0 & 2.7 & 0.0 & 2.1   \\
Best cand. selection & 5.5 & 0.0	& 1.5 & 0.0 & 1.5 &  0.0   \\
$\pi^\pm/K^{\pm}$ identif.  & 7.0 & 7.0	& 7.0 & 7.0 & 7.0 & 7.0  \\
$K_{S}$  reconstruction      & 1.1   & 1.1 & 1.1 & 1.1 & 1.1& 1.1  \\
 $\pi^{0}$ reconstruction & 1.1 & 1.1 & 1.1 & 1.1 & 1.1 & 1.1 \\
$\gamma$& - &-	& 3.8 & 3.8 & 7.6 & 7.6 \\
Tracking    &  2.2 & 2.2 & 2.2 & 2.2 & 2.2	& 2.2  \\
Polarization    &  0.0 & 0.0 & 0.8 & 2.8 & 0.6 & 0.2  \\
\hline
MC statistics for $\varepsilon$ & 0.2 &  0.2	& 0.4  & 0.4 &  0.5 &  0.5 \\
$D_{s}^{(*)}$ br.~fractions  & 8.6& 8.6& 8.6&8.6& 	8.7& 8.7 \\
$N_{B_s^{(*)}B_s^{(*)}}$            & \multicolumn{6}{c}{18.3} \\
$f^{}_{B^*_s\overline{B}^*_s}$ & \multicolumn{6}{c}{2.0} \\
\hline
Total  & 22.7 & 21.8  &  22.7 & 22.9 & 26.2 & 25.5 \\
\hline
\hline
\end{tabular*}
\end{table}

In the limits
$m^{}_{(b,c)}\ra\infty$ with $(m^{}_b-2m^{}_c)\ra 0$ and
$N^{}_c ({\rm number\ of\ colors})\ra\infty$, the 
$D^{*\pm}_s D^{\mp}_s$ and $D^{*+}_s D^{*-}_s$ modes
are \cp-even and (along with $D_s^+ D_s^-$)
saturate the width difference
$\Delta\Gamma^{CP}_s$~\cite{Aleksan}.
Assuming negligible \cp\ violation ($\phi^{}_{12}\!\approx\!0$), 
the branching fraction is related to $\dgs$ via
$\dgs/\gs = 2{\cal B}/(1-{\cal B})$. 
Inserting the total ${\cal B}$ from Table~\ref{tab:fit_results} 
gives $\dgs/\gs =  0.090\pm0.009 \,\pm0.023$, where the first 
error is statistical and the second is systematic. 
The central value is consistent with, but lower than, the 
theoretical prediction~\cite{Nierste}; the difference may be 
due to the unknown \cp-odd component in $B^0_s\ra D^{*+}_sD^{*-}_s$, 
and contributions from three-body final states. 
With more data these unknowns can be measured.
The former is estimated to be only 6\% for analogous $B^0\ra D^{*+}D^{*-}_s$
decays~\cite{Rosner}, but the latter can be significant: 
Ref.~\cite{Hou} calculates
$\Delta\Gamma(B^{}_s\ra D^{(*)}_s D^{(*)} K^{(*)})/\Gamma^{}_s=
0.064\pm 0.047$.
This calculation predicts $\Delta\Gamma^{}_s/\Gamma^{}_s$ from 
$D^{(*)+}_s D^{(*)-}_s$ alone to be $0.102\pm 0.030$, which 
agrees well with our result.
This agreement holds for $\phi_{12}$ values 
up to $\sim\!40^\circ$~\cite{phi12nonzero}. 

\begin{figure}[ptbh]
\hbox{
\hskip0.20in
\epsfig{file=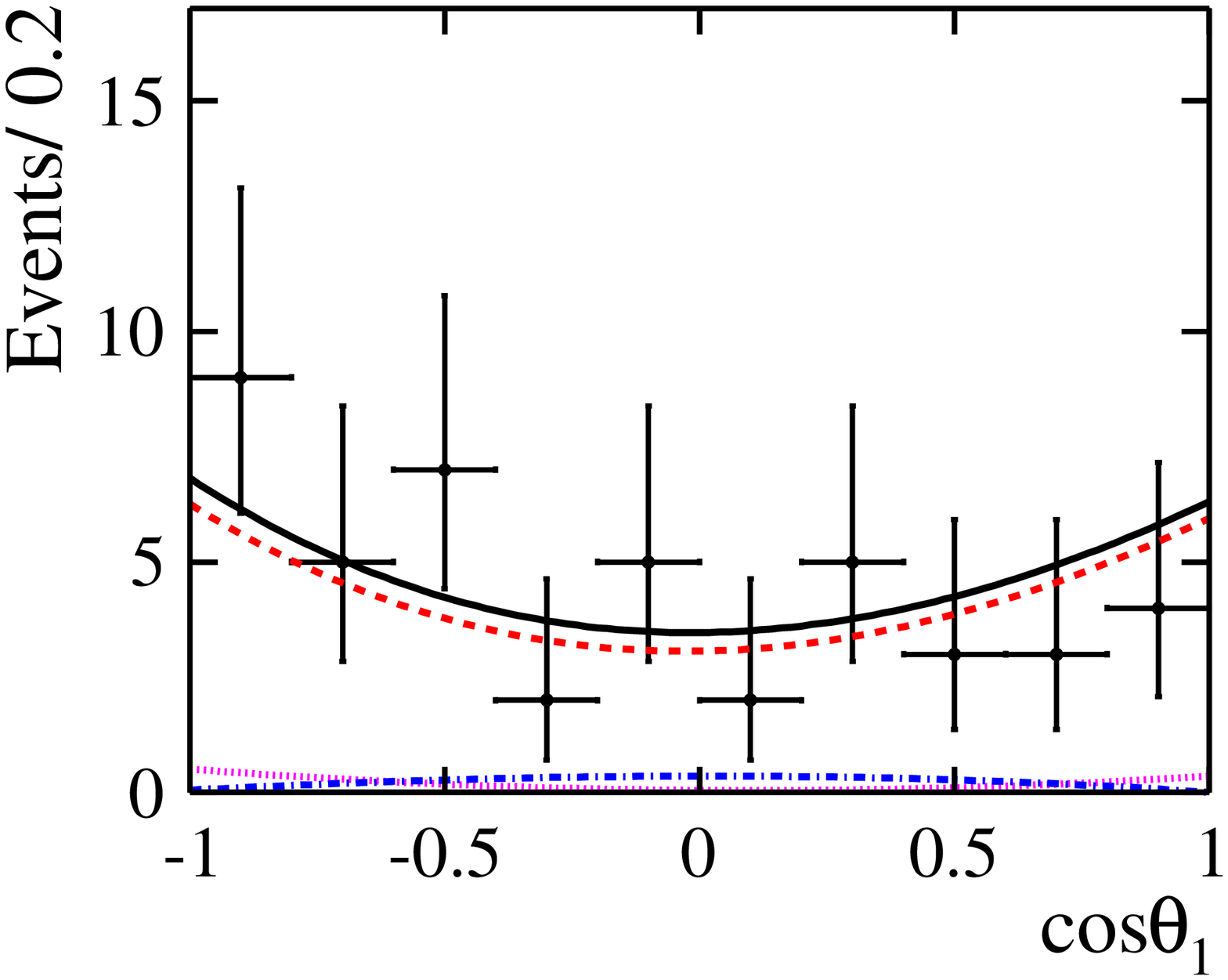,width=2.8in}
\hskip0.10in
\epsfig{file=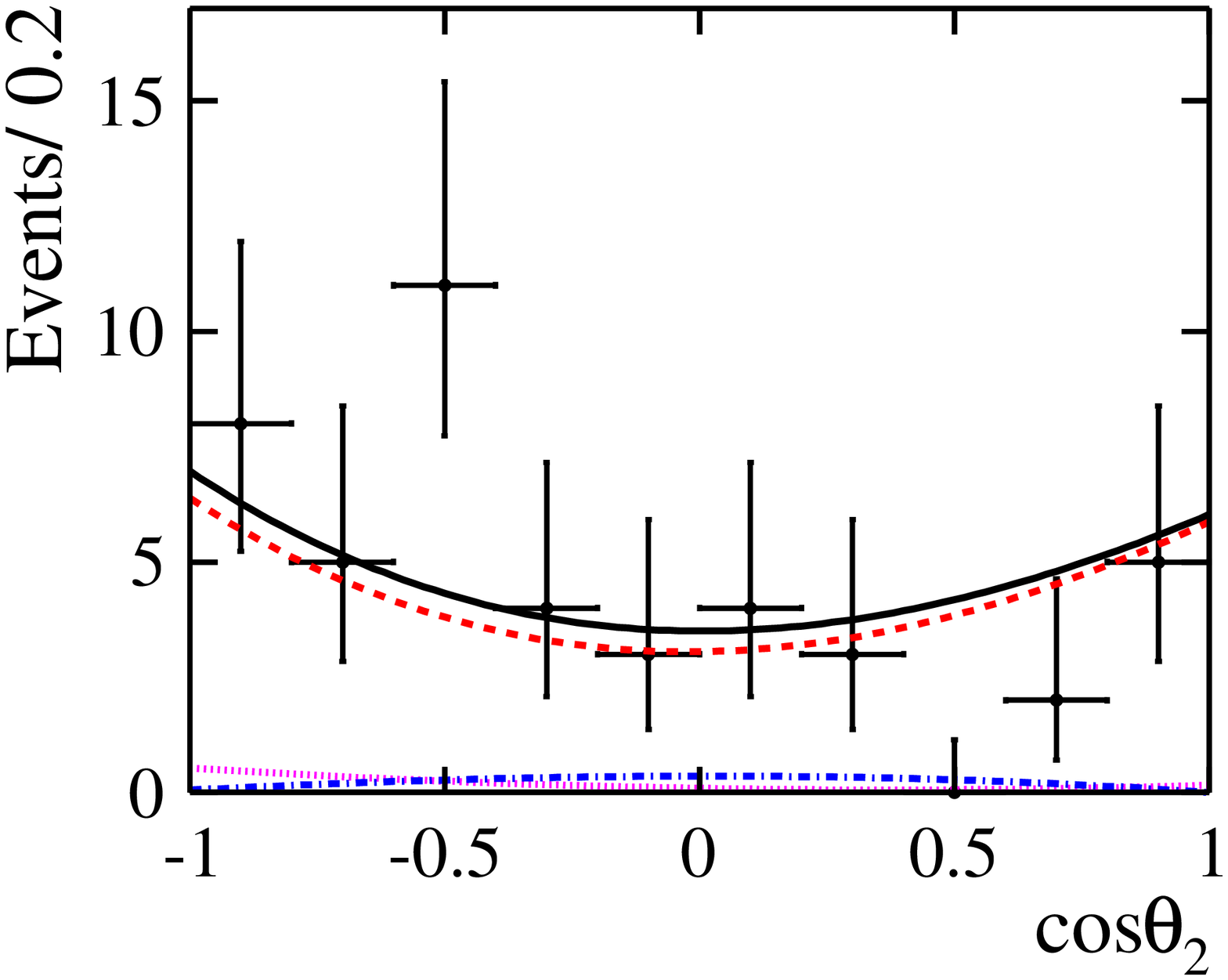,width=2.8in}
}
\vskip-0.15in
\caption{
Helicity angle distributions and projections of the fit 
result. The red dashed (blue dash-dotted) curves show 
the transverse (longitudinal) components; the magenta
dotted curves show background; and the black solid
curves show the total.
}
\label{fig:helicity_angles}
\end{figure}

To measure $f^{}_L$, we select events using the same criteria 
as before but, to minimize \bsdsstds\ cross-feed, we use a 
narrower range of $\mbc$ and $\de$ ($2.5\sigma$ in resolution). 
For these events we perform an unbinned ML fit to the helicity 
angles $\theta^{}_1$ and $\theta^{}_2$, which are the angles 
between the daughter $\gamma$ momentum and the opposite of 
the $B^{}_s$ momentum in the $D^{*+}_s$ and $D^{*-}_s$ rest 
frames, respectively. The angular distribution is
$\left(|A_+|^2 + |A_-|^2\right)\left(\cos^2\theta_1 +1\right)
\left(\cos^2\theta_2 +1\right) +
|A^{}_0|^2 4\sin^2\theta^{}_1\sin^2\theta^{}_2$,
where $A^{}_+$, $A^{}_-$, and $A^{}_0$ are the three
polarization amplitudes in the helicity basis.
The fraction $f^{}_L$ equals $|A_0|^2/(|A_0|^2+|A_+|^2 + |A_-|^2)$.
To account for resolution and efficiency variation, 
the signal PDFs are taken from MC. The background PDF 
is taken from an $\mbc$ sideband;
the level ($1.8\pm 0.7$ events) is estimated from a 
$D^+_s$ mass sideband and fixed in the fit. We obtain
\begin{eqnarray}
f^{}_L & = & 0.06\,^{+0.18}_{-0.17}\,\pm 0.03\,,
\label{eqn:helicity_result}
\end{eqnarray}
where the systematic errors arise from:
signal PDF shapes ($+0.008,-0.010$),
the background PDF shape ($+0.007,-0.004$),
fixed WC fractions ($+0.013,-0.015$),
the fixed background level ($\pm 0.022$),
$q\bar{q}$ suppression ($+0.011,-0$), 
possible fit bias ($+0, -0.011$), 
and MC efficiency due to statistics ($\pm 0.0004$).
The helicity angle distributions and fit projections
are shown in Fig.~\ref{fig:helicity_angles}.

In summary, we have measured the branching fractions for \bsdsds\ 
using $e^+e^-$ data taken at the $\Upsilon(5S)$ resonance. 
Under some theoretical assumptions and neglecting \cp\ violation,
the total branching fraction gives a constraint on~$\dgs/\gs$. 
We have also made the first measurement of the 
$B^0_s\ra D_s^{*+}D_s^{*-}$ longitudinal polarization fraction.

We thank R.\,Aleksan and L.\,Oliver for useful discussions.
We thank the KEKB group for excellent operation of the
accelerator; the KEK cryogenics group for efficient solenoid
operations; and the KEK computer group, the NII, and 
PNNL/EMSL for valuable computing and SINET4 network support.  
We acknowledge support from MEXT, JSPS and Nagoya's TLPRC (Japan);
ARC and DIISR (Australia); NSFC (China); MSMT (Czechia);
DST (India); INFN (Italy); MEST, NRF, GSDC of KISTI, and WCU (Korea); 
MNiSW (Poland); MES and RFAAE (Russia); ARRS (Slovenia); 
SNSF (Switzerland); NSC and MOE (Taiwan); and DOE and NSF (USA).


\begin{thebibliography}{99}
\bibitem{charge-conjugates} 
Charge-conjugate modes are implicitly included.
\bibitem{Novosibirsk} The Novosibirsk function is defined as
$f(x) = A\,\exp ( -0.5 
\{ \ln^2 [1 + \Lambda \tau \cdot (x - x_0) ] / \tau^2 + \tau^2 \} )$, 
where $\Lambda = 
\sinh ( \tau \sqrt{\ln 4} ) / ( \sigma \tau \sqrt{\ln 4} )$.
\end{thebibliography}
\end{document}